# Hydrogen Penetration and Fuel Cell Vehicle Deployment in the Carbon Constrained Future Energy System

Andrew Chapman[*,1], Dinh Hoa Nguyen[1], Hadi Farabi-Asl[2], Kenshi Itaoka[1], Katsuhiko Hirose[1], Yasumasa Fujii[3]
[1]Energy Analysis Division, International Institute for Carbon Neutral Energy Research, Fukuoka 819-0395, Japan
[2]Research Institute for Humanity and Nature (RIHN), Kyoto 603-8047, Japan
[3]Department of Nuclear Engineering and Management, University of Tokyo, Tokyo 113-8656, Japan
chapman@i2cner.kyushu-u.ac.jp, hoa.nd@i2cner.kyushu-u.ac.jp, farabi@chikyu.ac.jp, k.itaoka@i2cner.kyushu-u.ac.jp,
katsuhiko_hirose_aa@mail.toyota.co.jp, fujii@n.t.u-tokyo.ac.jp

*Abstract*— **This research details outcomes from a global model which estimates future hydrogen penetration into a carbon constrained energy system to the year 2050. Focusing on minimum and maximum penetration scenarios, an investigation of global fuel cell vehicle (FCV) deployment is undertaken, cognizant of optimal economic deployment at the global level and stakeholder preferences in a case study of Japan. The model is mathematically formulated as a very large-scale linear optimization problem, aiming to minimize system costs, including generation type, fuel costs, conversion costs, and carbon reduction costs, subject to the constraint of carbon dioxide reductions for each nation. Results show that between approximately 0.8% and 2% of global energy consumption needs can be met by hydrogen out to the year 2050, with city gas and transport emerging as significant use cases. Passenger FCVs and hydrogen buses account for almost all of the hydrogen-based transportation sector, leading to a global deployment of approximately 120 million FCVs by 2050. Hydrogen production is reliant on fossil fuels, and OECD nations are net importers – especially Japan with a 100% import case. To underpin hydrogen production from fossil fuels, carbon capture and storage (CCS) is required in significant quantities when anticipating a large fleet of FCVs. Stakeholder engagement suggests optimism toward FCV deployment while policy issues identified include necessity for large-scale future energy system investment and rapid technical and economic feasibility progress for renewable energy technologies and electrolyzers to achieve a hydrogen economy that is not reliant on fossil fuels.**

## I. Introduction

To contain global temperature rises to within 2 degrees Celsius of pre-industrial levels, the energy system must be rapidly decarbonized. If the more ambitious Paris Agreement target of 1.5 degree temperature constraint is to be met, the global energy system must be carbon neutral by the year 2050. Carbon reduction must be undertaken comprehensively, incorporating the electricity sector, transportation fuels and the use of complementary carbon dioxide ($CO_2$) removal technologies, while increasing overall electrification levels [1]. The Paris Agreement provides the mechanism for nations to reduce their $CO_2$ levels, known as National Determined Contributions (NDCs; [2]) which will allow the global energy system to reach a peak in carbon emissions in the short term, with medium to long term goals of rapid reduction in GHG levels, differentiated according to individual nations' ability to reduce emissions over time [3].

Moving away from the current fossil fuel dominated energy regime will not be simple, and will require a rapid transition toward alternative energy sources, prominent among them solar and wind power, reputed for their low marginal cost, however, having the issue of intermittent supply and a necessity for energy to overcome this issue and provide a stable, low carbon energy system [4]. Hydrogen is one option for renewable energy storage, and more broad uses such as vehicle fuels or chemical feedstocks. Hydrogen can assist in overcoming renewable energy intermittency issues, as part of the hydrogen economy. Hydrogen is not likely to be the single solution for the future low carbon energy system, but will be complemented by technologies such as carbon capture and storage (CCS) and emerging technologies such as Bioenergy with CCS (BECCS) and other negative emission technologies [5].

This research has three main aims: 1) to estimate the potential societal penetration of hydrogen across end uses, utilizing a global optimization-based model; 2) to demonstrate the impacts of hydrogen penetration on society, focusing on the transportation sector (i.e. FCVs and other vehicle types), and 3) to contrast model outcomes for the transportation sector with stakeholder preferences in the case study nation of Japan. This work represents a novel hybrid approach to transportation system modelling cognizant of both optimized model outcomes and stakeholder preferences.

The remainder of this paper is divided into the following sections: Section II deals with the background, detailing previous scholarship on hydrogen deployment for transport, storage and industry at the regional and national level, while identifying the gaps which are addressed here. Section III outlines the methodology used, for the optimization model, stakeholder engagement and focus on fuel cell vehicle deployment analysis. Section IV details the results for the global energy system for 2000 to 2050, including hydrogen production sources and end uses, with a strong focus on the transport sector, system cost and carbon reduction issues alongside stakeholder preferences for future automobile purchase. Section V discusses the findings, bringing together model outcomes, policy issues and future challenges. Section VI outlines the conclusions.

## II. Background

Previous research has investigated the penetration of hydrogen into the future energy system, predominantly by end use or by specific sector. This section details pertinent research from the transport and industrial sector, storage and regional studies, clarifying the novelty of the present study.

The transport sector has been rigorously studied, considering electric and hydrogen fueled vehicles from an economic perspective, and in terms of an optimized transport sector [6-8]. The potential of electronic vehicles as an alternative to hydrogen vehicles, as discussed in this study, demonstrated their potential positive impact on decreasing system costs, particularly when renewable energy is prevalent [9]. A Malaysian study on the role of hydrogen in the future transport system was highlighted as a long-term sustainable option [10]. In addition, a study of FCVs in New Zealand highlighted their potential to fully displace conventional gasoline vehicles, while also highlighting the challenges remaining in terms of motor design, system configuration and optimization techniques, which when overcome could elevate FCVs to a promising alternative worldwide [11]. Research which optimizes FCVs according to factors such as traction, hydrogen use and energy recovery bode well for future FCV development [12], particularly considering stakeholders preferences for energy efficiency as identified in this study. In terms of transportation relevant to mass transit and freight, a study on the potential conversion of railway vehicles to a hydrogen or hydrogen-hybrid type demonstrated the potential for a 44-60% reduction in energy consumption and a commensurate reduction in $CO_2$ emissions of 59-77% [13].

In addition, hydrogen use in the industrial sector has been evaluated, focusing on hydrogen as a chemical feedstock, specifically in terms of syngas, methanol, ammonia and for the production of hydrocarbons utilizing the Fischer-Tropsch process [14] and theoretical limits of hydrogen blending with city gas, identifying a threshold of 30% without the need for modified infrastructure [15]. One challenge for hydrogen production into the future is the shift away from predominantly fossil fuel based production toward more sustainable approaches [16].

In terms of storage, hydrogen shows promise as a seasonal scale storage medium, ameliorating renewable energy intermittency and encouraging additional deployment [17]. In terms of cost however, battery banks were found to be more economically viable when considered for augmentation of the current energy system [18]. In terms of optimizing renewable energy deployment cognizant of the potential of power to gas arrangements, wind power emerges as a potential beneficiary of such a system, engendering lower levels of curtailment and reduced emissions [19]. This finding is supported by separate research which identifies surplus renewable energy (particularly wind; [20]) to hydrogen as a potential mechanism for environmentally friendly hydrogen as both a fuel and storage medium [18].

Precedential research has been undertaken at the regional level, either within individual nations or across grid-connected regions, i.e. those found within Asia or Europe. The current study uses an established optimization model known as Dynamic New Earth (DNE) which is able to consider the energy system at the global level, not just for interconnected regions but also regions separated by ocean, reliant on maritime energy trade. This model takes into account future carbon reduction goals, system costs, including those for energy resources and the need for additional infrastructure deployment. On the supply the model is cognizant of energy generation capacity factors and efficiencies, while on the demand side energy use and conversion and imports and exports between regions are also accounted for [21]. This study relies on the modelling approach seeking an optimal future energy system, cognizant of carbon targets, energy policy and technology learning curves for the provision of minimum and maximum hydrogen penetration scenarios from 2000 to 2050 [22].

This research addresses a significant gap in existing work, and details a future energy system out to 2050 cognizant of the Paris Agreement and regionally differentiated carbon reduction targets. The outputs of the model identify the quantity of hydrogen produced, the fuels and regions responsible for its production, as well as the final use cases. The transport sector is the main focus of this research, and the number, type, and region of deployment of vehicles (particularly FCVs) is discussed in detail. This novel research allows for a consideration of optimized energy systems from a cost and technology perspective, cognizant of learning curves and emerging technologies, while being responsive to stakeholder preferences for energy use, the environment, affordability and convenience, particularly relevant to the emergence of FCVs as a personal transportation option.

## III. Methodology

The methodology for this research is conducted in two parts. First, the global energy model is described, including the detailed assessment of fuel cell and other vehicle deployment. Second, stakeholder engagement is undertaken in Japan, a nation in which the hydrogen economy is being aggressively pursued, to determine stakeholder preferences and reasoning for vehicle purchase, now and in the future, providing a contrast with global and regional model results.

### A. Global Model

The global model aims to estimate the quantity of hydrogen introduced to the energy system between 2000 and 2050, detailing the source and region of production and consumption. Focusing on (fuel cell) vehicle deployment, hydrogen-based use cases are quantified. To achieve these goals, a global linear optimization model is utilized according to the following assumptions, policy and technology constraints:

1) The global model accounts for 82 regions, representative of population centers and resource extraction characteristics.

2) Interaction between regions is cognizant of grid connections and international pipelines, as well as the need for maritime trading for regions which are not connected.

3) End uses for hydrogen span across transport (i.e. FCVs), city gas blending up to maximum theoretical values [15], as well as electricty generation either through direct combustion or co- firing with natural gas, and use as a chemical feedstock (e.g. methanol synthesis).

4) Carbon reduction targets out to 2050 are utilized as an input for the model, relying on the Representative Pathways 2.6 (RCP 2.6; [23]), which aligns with Paris Agreement 2-degree targets, considering both equal and differentiated OECD and non-OECD nation contributions towards these goals.

5) Learning curves and economic outlooks for renewable energy generation and carbon reduction technologies are incorporated [24,25], including a specific focus on vehicle costs and learning curves [26,27].

6) Mimum and maximum investigated scenarios consider both restricted nuclear deployment cognizant of the recent nuclear accident in Japan, World Nuclear Association estimates for national deployment levels [28] as well as an unrestricted nuclear deployment scenario.

7) CCS is used, with priority given to storage in depleted gas wells and saline aquifers. $CO_2$ which is not sequestered is made available for pipeline, land or maritime transport for use in enhanced coal bed methane or enhanced oil recovery where econmically feasible.

The global linear optimization model is described generally below:

minimize
$$\sum_{t=1}^{T}\sum_{n=1}^{N}\sum_{i,j,l,k,m,p} g\left(x_i^f(t), x_j^r(t), x_l^n(t), y_k^{lf}(t), y_m^{gf}(t), s_h(t)\right)$$

subject to:
$$\sum_{i,j,l,k,m,p} x_i^f(t) + x_j^r(t) + x_l^n(t) + y_k^{lf}(t) + y_m^{gf}(t) + s_h(t) = d(t)$$
$$C_n(T) \leq cap_n(T), n = 1, \dots, N$$

where
$$g\left(x_i^f(t), x_j^r(t), x_l^n(t), y_k^{lf}(t), y_m^{gf}(t), s_h(t)\right)$$
$$= p_i^f(t)x_i^f(t) + p_j^r(t)x_j^r(t) + p_l^n(t)x_l^n(t)$$
$$+ p_k^{lf}(t)y_k^{lf}(t) + p_m^{gf}(t)y_m^{gf}(t)$$
$$+ p_h(t)s_h(t)$$
is the total system cost at time step t.

Variables and parameters contained in the model are described in Table 1.

Table 1. Variables and parameters used in the global model.

| Variable | Definition |
|---|---|
| $t, T$ | Time step, number of time steps |
| $n, N$ | Country index, number of countries |
| $i, j, l, k, m$ | Indexes of different energy resources |
| $f$ | Fossil fuels (coal, gas, oil, etc.) |
| $r$ | Renewables (solar, wind, hydro, etc.) |
| $n$ | Nuclear (LWR, FBR, HTGR, etc.) |
| $lf$ | Liquid fuels (gasoline, methanol, etc.) |
| $gf$ | Gaseous fuels, (hydrogen, methane, etc.) |
| $x_i^f(t)$ | Amount of fossil fuel type $i$ at time step $t$ |
| $x_j^r(t)$ | Amount of renewable type j at time step $t$ |
| $x_l^n(t)$ | Amount of nuclear type $l$ at time step $t$ |
| $y_k^{lf}(t)$ | Amount of liquid fuel type $l$ at time step $t$ |
| $y_m^{gf}(t)$ | Amount of gaseous fuel type $l$ at time step $t$ |
| $s_h(t)$ | Storage type $h$ at time step $t$ |
| $p_i^f(t)$ | Price of fossil fuel type $i$ at time step $t$ |
| $p_j^r(t)$ | Price of renewable type $j$ at time step $t$ |
| $p_l^n(t)$ | Price of nuclear type $l$ at time step $t$ |
| $p_k^{lf}$ | Price of liquid fuel type $k$ at time step $t$ |
| $p_m^{gf}(t)$ | Price of gaseous fuel type $m$ at time step $t$ |
| $p_h(t)$ | Price of storage type $h$ at time step $t$ |
| $d(t)$ | Final demand at time step $t$ |
| $C_n(T)$ | Amount of carbon emission in country $n$ at final time step $T$ |
| $cap_n(T)$ | Upper limit of carbon emissions in country $n$ at final time step $T$, calculated by its given carbon reduction target |

The model optimizes the energy system in terms of total system cost and carbon budgets in 10 year time slices, and is cognizant of the global energy system from both the supply and demand side. Due to the large scale optimization problem engendered by the number of nations, fuel types, prices and technologies considered over time, IBM ILOG CPLEX Optimization Studio is employed in this research.

The primary energy sources accounted for include fossil fuels, i.e. coal, oil, natural gas, and a group of unconventional fuels, including heavy crude oil, oil sands and shale oil. Nuclear energy types include light water and light water mixed oxide (MOX) fueled reactors as well as fast breeder reactors (FBR) and high temperature gas cooled reactors (HTGR) including MOX varieties. For renewable energy, the model is cognizant of solar photovoltaics (PV), wind, geothermal, hydropower and biomass (including specific biomass crops, agricultural and industrial residues, bagasse and household wastes).

Secondary energy carriers, while focusing on hydrogen, also include methane, methanol, dimethyl ether (DME), oil products, carbon monoxide and electricity. Final energy demands are met using solid, liquid and gaseous fuels and electricity, meeting daily energy demands in line with seasonal fluctuations. In assessing energy system costs, in addition to the primary, secondary and final energy demands, raw material costs, transportation costs and production costs for each region, cognizant of losses are calculated.

The impacts and feasibility of CCS and energy storage media (including pumped hydro, battery storage and compressed air storage) are assessed regionally, accounting for cost and deployment limits [22].

The focus of this study is the transportation sector, specifically the potential penetration of FCVs into the future energy system. The assumptions used for vehicle costs refer to projections in price differentials between conventional gasoline hybrid vehicles and FCVs [29].

*B. Stakeholder Engagement and Case Study*

Stakeholder engagement is undertaken considering two key aspects affecting vehicle deployment; perception of automobile types from the aspects of energy efficiency, environment and accessibility to refueling infrastructure and affordability, and, opinions on the types of cars people are likely to buy, now and 10 years in the future.

The survey of perception and purchase choice was undertaken via internet survey in Japan in February 2019. The representative sample size was 5,000, including people over the age of 20 with a driver's license, balanced across region, gender and age group.

Perception was measured toward the factors of energy efficiency, environment, accessibility and affordability on a 4 point scale; 'good', 'moderate', 'bad' and 'don't know'. Purchase choice was measured across the three options of 'would consider buying', 'Would need more research before considering buying', and 'would not consider buying'. A follow up question was asked using the same scale for 10 years in the future, under the assumption that refueling infrastructure had progressed considerably.

The research design allows for both a quantitative evaluation of global energy system outcomes under varying energy scenarios and a qualitative evaluation of automobile perceptions and purchasing choices in an energy system transitioning to low carbon options including renewable energy and hydrogen. Fig. 1 details the quantitative model flows and focus of this study, along with linkages to the qualitative investigation in Japan focusing on automotive applications and stakeholder engagement.

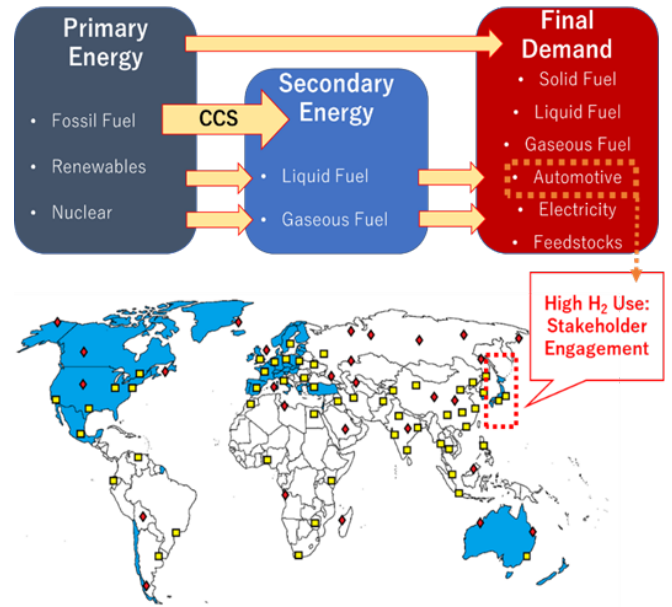

Fig. 1. Model nodes, primary and secondary energy types, final demand including automotive applications and stakeholder engagement in Japan

IV. RESULTS

Results of this study are divided into four sub-sections for ease of understanding, beginning with 1) Global energy system composition under maximum and minimum hydrogen penetration scenarios, 2) Resultant hydrogen production sources and consumption end uses, focusing on (fuel cell) vehicle deployment, 3) Energy system economic and contributions from alternative carbon reduction technologies, and 4) Stakeholder perceptions and purchase choice outcomes.

*A. Global Energy System Composition Under Maximum and Minimum Hydrogen Penetration Scenarios*

The two scenarios presented in this research are the maximum and minimum hydrogen penetration scenarios, as detailed in [22]. For the maximum hydrogen penetration scenario, city gas blend level is set to a 30% $H_2$/City Gas ratio, nuclear power deployment is restricted in line with national policies and carbon targets are differentiated between OECD (80% $CO_2$ reduction required by 2050) and non-OECD (60% by 2050) nations, representing differentiated carbon reduction goals, according to development level [3]. For the minimum hydrogen penetration scenario, the city gas blend level is reduced to a conservative 5% level, nuclear power deployment is unrestricted and carbon targets are identical for OECD and non-OECD nations, set to a 65% reduction of $CO_2$ by 2050.

Between the year 2000 and 2050, we note a shift in energy sources, notable a strong move away from fossil fuels (i.e. coal, oil and natural gas), in preference for low-carbon generation options such as nuclear power (LWR, MOX, FBR and HTGR, which can produce hydrogen without the need for electrolysis) and renewables, which are dominated by hydro, wind, solar and biomass technologies.

For the maximum hydrogen penetration case (Fig 2a.) FBR and HTGR technologies do not emerge due to restrictions on nuclear power deployment, while for the minimum hydrogen penetration case (Fig 2b.) we note the emergence of FBR nuclear power from 2030 and LWR-MOX and HTGR technologies from 2040 onwards.

Although Fig. 2 details a significant overall increase in nuclear power generation out to 2050, this is predominantly observed in non-OECD nations, while for OECD nations the total contribution from nuclear power in 2050 is only approximately 1% of power generation for the maximum penetration case (Fig. 2a), while nuclear power's contribution to overall generation is increased for all nations under the minimum hydrogen penetration, unrestricted nuclear scenario (Fig. 2b).

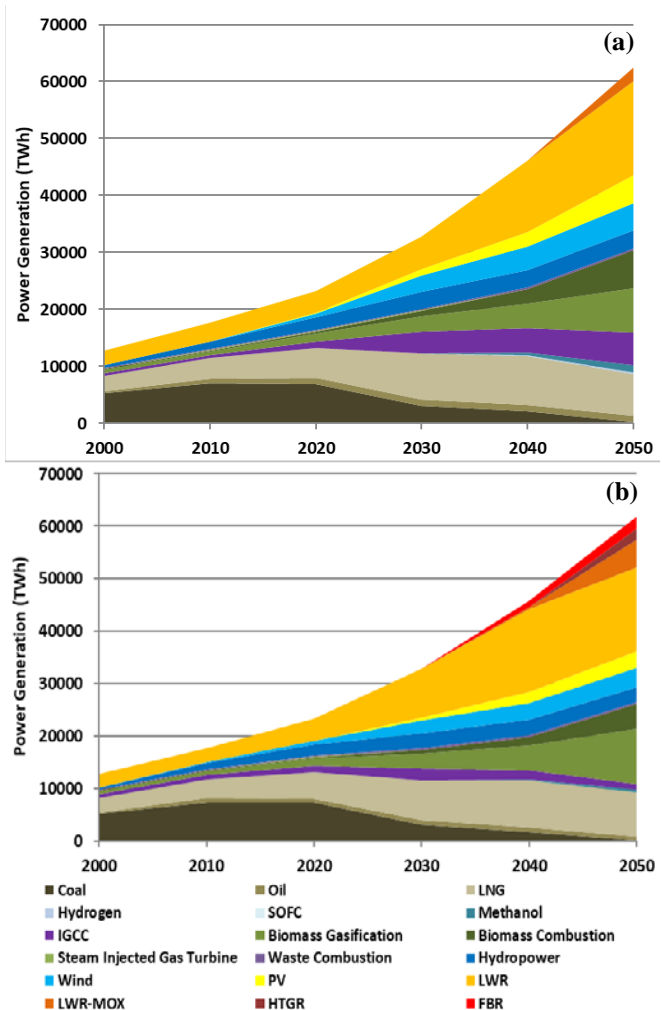

Fig. 2. Model results for power generation sources from 2000-2050 for (a) maximum hydrogen, and (b) minimum hydrogen penetration cases

### B. Hydrogen Production Sources and End Uses

It is also possible to analyze in detail the source and end uses of hydrogen. For the maximum hydrogen penetration scenario, we note that the majority of hydrogen comes from the fossil fuel sources of natural gas, coal and oil, with a moderate contribution from biomass (Fig. 3a). For the minimum scenario, nuclear energy underpins the future hydrogen scenario (Fig. 3b).

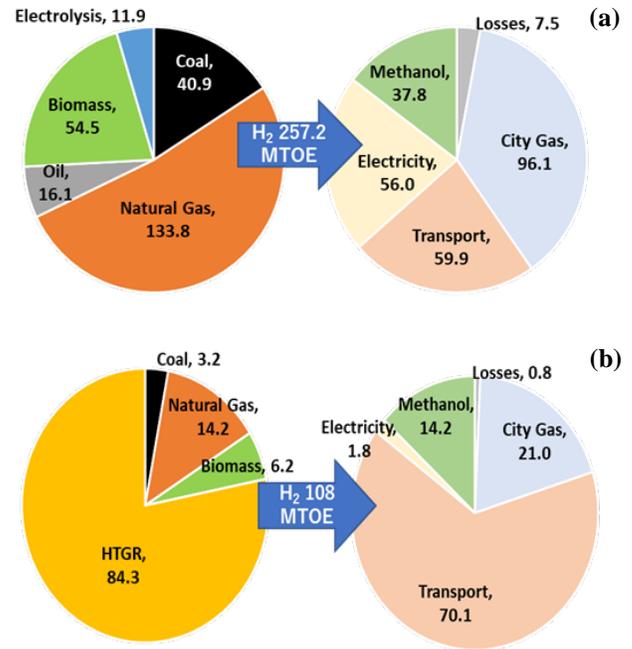

Fig. 3. Hydrogen production sources and use cases for the year 2050 as determined by the global model for (a) maximum and (b) minimum penetration scenarios

For the maximum hydrogen penetration case, hydrogen is produced in both OECD (170.9 megatons of oil equivalent (MTOE)) and non-OECD (86.3 MTOE) nations, however the vast majority (approximately 87%) is used in OECD nations. For the minimum penetration scenario the production and consumption of hydrogen is more even between the two regions, with a share of 35%/65% for production and 42%/58% for consumption for OECD and non-OECD nations respectively.

Considering both scenarios, Japan is a special case with very high hydrogen use in their future energy system. For the maximum penetration case, in 2050 Japan consumes approximately 41% of global hydrogen (106 MTOE), relying exclusively on imports. For the minimum scenario, a total of 17 MTOE of hydrogen (~16% of global hydrogen) is consumed in Japan, exclusively coming from HTGR nuclear reactors.

As shown in Fig. 3, the transport sector is a preferential use case in both the maximum and minimum scenarios in 2050. A key focus of this research is the makeup of the transport sector and the types of vehicles deployed in each region. For the maximum penetration scenario, transport accounts for 59.9 MTOE, made up of 33.1 MTOE for passenger vehicles, 15.2 MTOE for hydrogen buses and 11.6 MTOE for hydrogen based freight applications. Under the minimum hydrogen penetration scenario, transport accounts for the majority of hydrogen usage (due to direct derivation of electricity from nuclear power) and use cases are again dominated by passenger vehicles (46.4 MTOE) and buses (15.2 MTOE), followed by freight applications (8.5 MTOE). It is of note that the bus use case appears to be saturated in both scenarios.

In terms of passenger transportation, by 2050, the previously gasoline and gasoline hybrid dominated world markets shift to a market breakdown of gasoline hybrids, passenger FCVs and FCV buses in the ratio of 85.3%/11.1%/3.6% for the minimum penetration scenario and 88.8%/7.7%/3.5% for the maximum penetration scenario. Passenger FCVs emerge from 2040, the same year in which gasoline passenger vehicles are phased out.

*C.  System Costs and Complementary Carbon Reduction Technologies*

The estimate contribution of hydrogen to the future energy system in this research reaches a maximum level of 2% by 2050. This estimate is considerably smaller than the ambitious 18% penetration goal set by the Hydrogen Council [30].

To achieve the maximum estimated 2% hydrogen contribution to the future energy system while achieving Paris Agreement carbon reduction targets has significant implications for system costs and CCS deployment requirements. The overall energy system cost (i.e. the sum of fixed and variable costs) increases in excess of six times over the investigated period (Fig. 4). These increases are largely due to the necessity for expensive carbon reduction options such as CCS to meet carbon reduction targets. From an economic standpoint, not all cost increases are considered negative, and some nations will benefit in terms of new employment opportunities emerging from additional industrial activity.

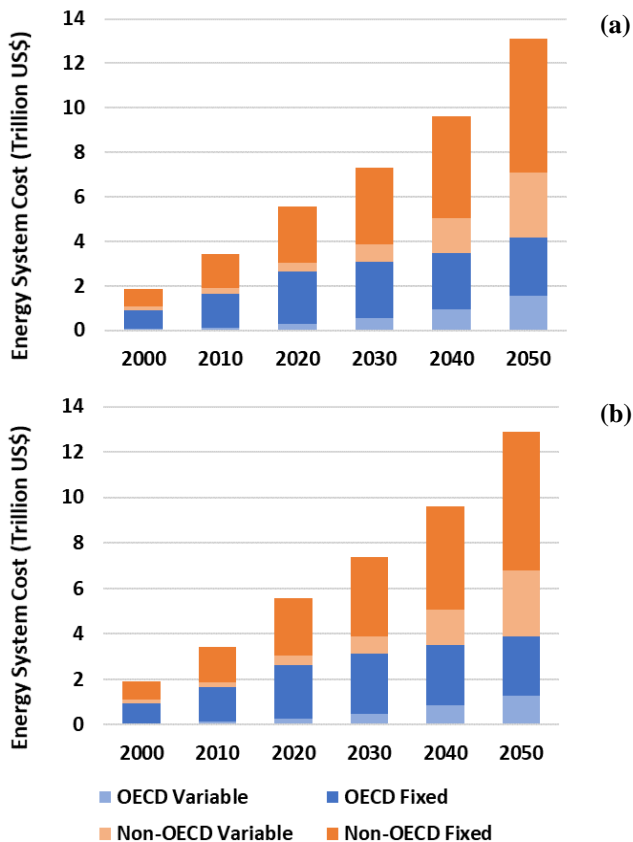

Fig. 4. Energy system costs (Trillion US$) 2000-2050 for the (a) Maximum, and (b) Minimum hydrogen penetration scenarios

The maximum hydrogen penetration scenario is the most expensive overall, growing to 13.1 trillion US dollars by 2050, compared to 12.9 trillion us dollars for the minimum penetration scenario. For both scenarios between 2000 and 2050, the non-OECD system costs grow by approximately nine times, compared to approximately four times for the OECD system costs, demonstrating the rapid growth of the non-OECD nations' energy system. A large portion of these costs is for CCS. The total deployment of CCS required to meet carbon reduction goals by 2050 is required to account for approximately 53% of total emissions (3902 megatons of $CO_2$) for the minimum and approximately 58.4% of total emissions (4931 megatons of $CO_2$) for the maximum hydrogen penetration scenario.

*D.  Stakeholder Engagement Outcomes*

In addition to quantitative energy system model results, building on the optimized future energy system deployment estimate for vehicles, stakeholder input provides additional evidence for a changing energy system. In this case we focus on Japan, as the major world user of hydrogen in our maximum penetration scenario, dependent on current world policies and carbon targets.

The first aspect of stakeholder engagement; vehicle perception and the likelihood of purchase is detailed in Fig. 5 for the six prominent vehicle types in Japan; Gasoline, Hybrid, Plug-in Hybrid, Diesel, Battery Electric Vehicle (BEV) and Fuel Cell Vehicle (FCV).

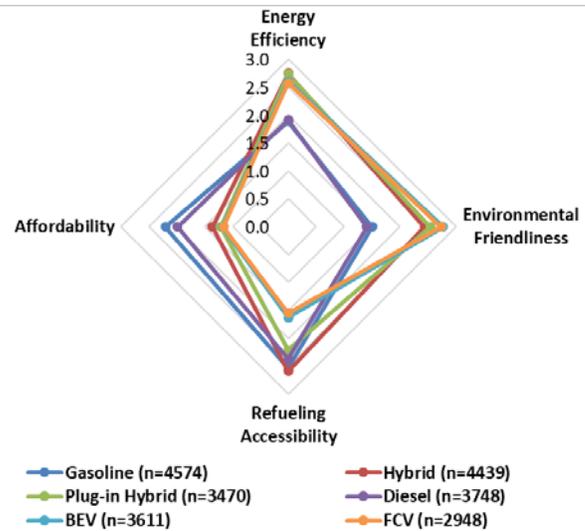

Fig. 5. Japanese consumer perceptions of passenger vehicles by type

According to stakeholders, gasoline and diesel vehicles are considered among the most affordable and accessible, yet the least environmentally friendly or efficient among the six investigated vehicle types. Hybrid vehicles were considered less affordable yet scored higher in terms of accessibility, environmental friendliness and energy efficiency. BEV, plug-in hybrid and FCV's scored similarly high scores for efficiency, environmental friendliness, although they were considered the least affordable among types. While plug-in hybrids scored relatively highly in terms of accessibility to refueling (likely due to being able to use both electricity and gasoline), this was not the case for FCV or BEV's. Considering the number of 'don't know' responses received, we can ascertain that stakeholders' familiarity with BEV's, plug-in hybrids and FCV's was lower than for other types, with FCV's the lease well understood among stakeholders..

Purchase likelihood, incorporating stakeholder views on vehicle properties is detailed in Fig. 6, detailing current views in panel a, and future views (circa 2030) in panel b.

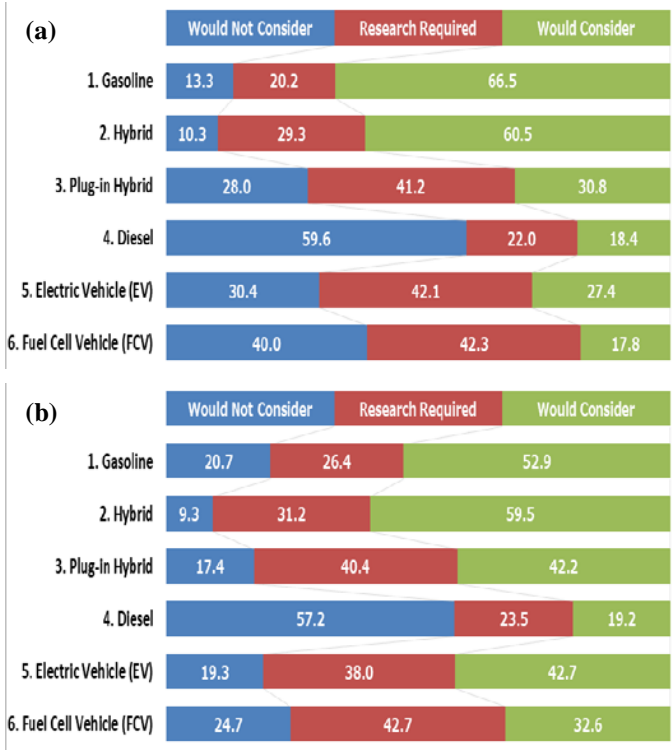

Fig. 6. Japanese consumer preferences for likelihood of vehicle purchase for (a) current vehicle performance and refuelling infrastructure, and (b) 2030, following improvements in vehicle performance and refuelling infrastructure

For current purchase preferences, the most popular choices are dominated by gasoline and hybrid vehicles. Followed by plug-in hybrids, BEV's, diesel vehicles and FCV's. Of the vehicles which stakeholders would not consider in their purchase choice, diesel was most strongly opposed followed by FCV's and BEV's (Fig. 6a). Considering their purchase preference in 10 years' time, under the assumption that refueling issues had been resolved and vehicle performance improved, we note that BEV and FCV's improve markedly as a purchase option. BEV's become preferable to plug-in hybrids, and FCV's improve their appeal overall, with less people preferring gasoline vehicles under these assumptions (Fig. 6b).

V. DISCUSSION AND IMPLICATIONS

In our global investigation, hydrogen has the potential to account for between 0.8% and 2% of global energy consumption, however within this quantity, hydrogen is particularly prominent within the transportation sector.

Focusing on the maximum penetration scenario, globally, hydrogen is estimated to underpin approximately 7.5% of transport sector energy consumption by 2050. Within the passenger vehicle market, hydrogen will provide approximately 13.5% of passenger kilometers by 2050 (the remaining 86.5% is made up predominantly of gasoline and gasoline-hybrid passenger cars). This level of hydrogen penetration of the transports sector leads to a the replacement of gasoline and hybrid passenger vehicles with approximately 120 million hydrogen passenger cars (deployment rates vary, depending on the region).

The deployment of passenger FCVs occurs mostly within OECD nations, which are generally richer and more developed than their non-OECD counterparts. Fig. 7 demonstrates this result, showing that OECD nations make up approximately 74% of the anticipated deployment by 2050, dominated by the strong hydrogen use case nation of Japan, North America and Western European nations. Non-OECD nations are represented predominantly by China and South East Asia. Building on the hydrogen penetration result, it is possible to estimate regional vehicle deployment numbers for the year 2050, taking into account average vehicle occupancy and annual travel distances for each region [31].

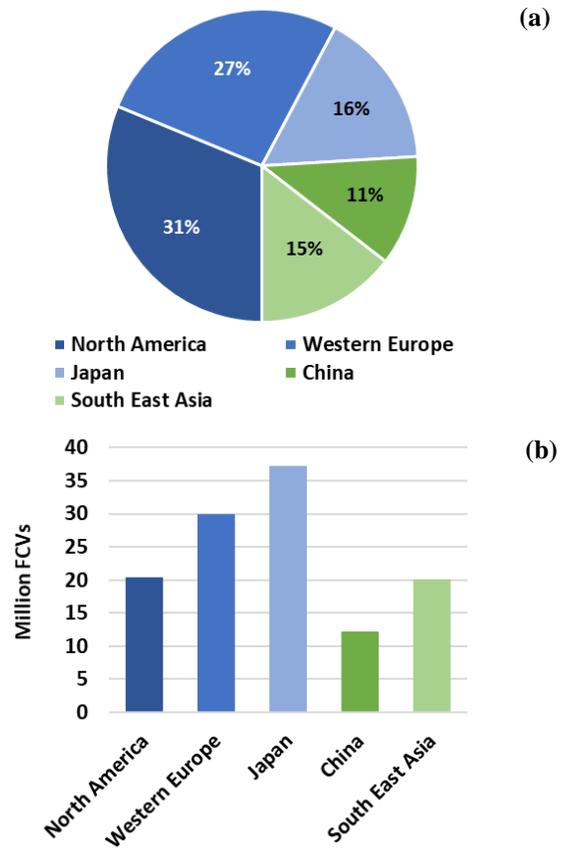

Fig. 7. (a) Regional hydrogen distribution (% of total global MTOE) for transportation, and (b) FCV regional deployment numbers in 2050

Although Japan's share of the global hydrogen transportation quanta is lower than that of other regions, due to the relatively low amount of kilometers travelled by each vehicle (approximately one-third of their US counterparts), the overall number of FCVs is highest in Japan by 2050, approximately 37.3 million vehicles. Currently in Japan, approximately 78.1 million passenger vehicles are in use, meaning that hydrogen will play a decidedly prominent role by the year 2050. For other

OECD nations, Western Europe deploys approximately 30 million, and North America 20.4 million FCVs by 2050. For non-OECD nations, the South East Asia region deploys approximately 20.2 million, while China deploys approximately 12.2 million FCVs by 2050.

Advances in vehicle technology means that FCVs are already in the market, and are rapidly improving their efficiency and affordability [32]. Currently, there are approximately 11,200 FCVs in operation globally, with projection suggesting an increase to approximately 2.5 million vehicles within the next ten years [33]. By 2050, we anticipate in excess of 120 million FCVs in the market, and this level of deployment will require sustained and increased deployment rates post-2030, reflective of improving affordability and competitiveness with traditional vehicle types.

To enable such a large deployment of FCVs in the future transport sector, additional investment in hydrogen infrastructure is required, particularly with regard to storage, distribution and refueling infrastructure. In addition, hydrogen importing nations (i.e. Japan) will need to construct and maintain a maritime import-export network. Fig. 3a, which detailed the maximum penetration scenario showed that the majority of hydrogen is being produced from fossil fuel sources, however Japan, which is likely to import all of their future hydrogen may be able to offset their own $CO_2$ emissions and realize a low carbon transport sector, while producer nations realize economic benefits at the expense of the environment.

The balance of hydrogen production and consumption is also of note. While non-OECD nations produce approximately one-third of global hydrogen needs in 2050, they consume only one-tenth, exporting 69% of this to OECD nations, underpinning a significant portion of the FCV deployment estimated in this study.

In addition to the identified strong role for hydrogen in terms of passenger cars, by the year 2050 hydrogen also contributes moderately to the mass transportation sector, leading to the deployment of FCV buses. FCV passenger car and bus efficiencies and costs are derived from publicly available fuel cell data from Toyota Motor Corporation (as used in the Mirai FCV).

Taking into account Japanese stakeholder preferences toward future car purchases, we observe a growing trend toward consideration of FCV vehicle purchase, to approximately one-third of stakeholders by 2030, compared to less than one-fifth of stakeholders in 2019. Should this growing acceptance of FCV's be maintained, and promised infrastructure and performance improvements eventuate, a fleet consisting of approximately 50% FCV's by 2050 in Japan is an ambitious, yet not unreasonable projection, especially considering complementary research outcomes for other regions [10,11]. Hybrid vehicle preferences remain strong throughout the tested time periods in Japan, and the assumption that gasoline-hybrid vehicles will continue to play a strong role, at the expense of gasoline vehicles appears sound.

Although the model outcomes represent an optimized energy system in terms of total system cost and achieving the Paris Agreement $CO_2$ reduction targets, there are some complementary outcomes in terms of stakeholder preferences. We note in the results (Fig. 5) that stakeholders are very optimistic about the environmental and energy efficiency aspects of FCV's. These perceptions are borne out in comparative studies of FCV and internal combustion engine vehicles, noting that overall FCV's maintain an advantage in terms of $CO_2$ per km and overall energy efficiency [34]. Due to these advantages, and in line with anticipated learning curves for FCVs, by 2050, it is predicted that that FCV affordability will be superior to that of gasoline vehicles, perhaps allaying the concerns expressed by stakeholders with regard to FCV cost.

When considering the minimum hydrogen penetration scenario, we note that although the quantity of hydrogen allocated to transport increases overall, the global energy system shifts to a local production and consumption approach – largely due to non-preferential carbon reduction targets for non-OECD nations. Under this arrangement, even Japan produces all of its own hydrogen from HTGR nuclear facilities, however, a much smaller quantity is therefore used in Japan (17.1 MTOE vs 106 MTOE for the maximum scenario). The non-preferential carbon reduction targets mean that import does not occur, and no hydrogen is allocated to passenger transportation, except for FCV buses at the same level as the maximum scenario (see Fig. 8).

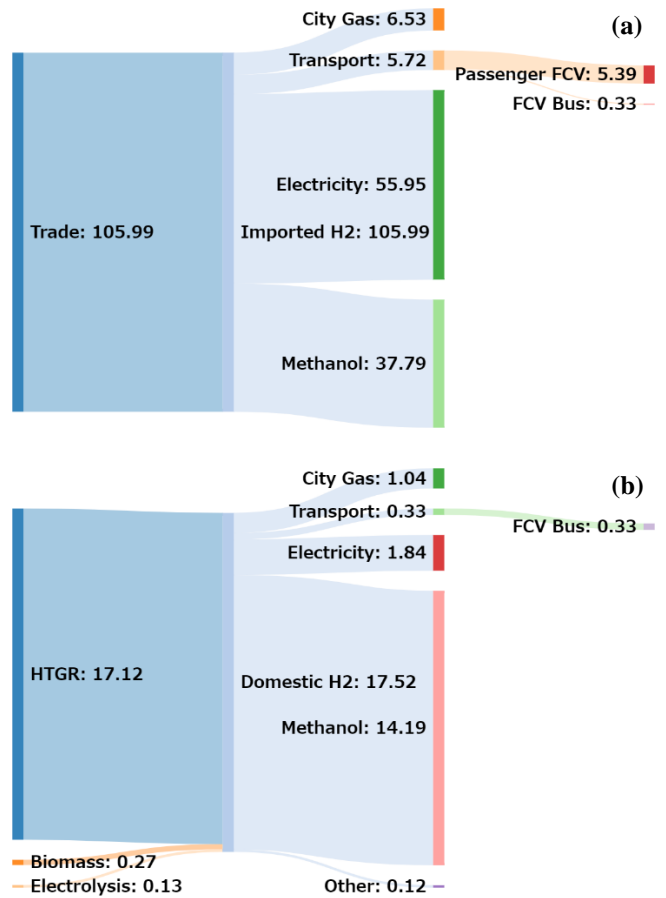

Fig. 8. Japanese hydrogen penetration scenarios (a) Maximum penetration reliant on imports, and (b) Minimum penetration, underpinned by domestic supply. Units are in MTOE.

Considering both scenarios, of outstanding concern is the requirements for CCS in order to meet $CO_2$ reduction targets. Cognizant of potential stakeholder concern toward CCS projects, and the long period required to commercialize pilot-scale projects, it is uncertain that by 2050 $CO_2$ reduction targets can be met utilizing the current suite of technologies and CCS alone. In order to reduce the reliance on CCS noted in the model outcomes, significant progress is required for alternative technologies including electrolyzers and renewable energy generation. Should this suite of technologies not be able to meet emission reduction targets, emerging technologies which offer negative emissions such as Bio-Energy with Carbon Capture and Storage (BECCS) or direct air capture (DAC) technologies may also need to be incorporated.

## VI. Conclusions

This research elucidates a significant role for hydrogen in the future energy system, focusing on the transport sector, yet cognizant of a range of sectors including city gas, chemical feedstocks and direct firing for electricity generation. At the global scale, we estimate that hydrogen will account for between approximately 0.8% and 2% of global energy consumption, with the majority consumed in OECD nations. For the maximum penetration scenario, a city gas blend incorporating the theoretical limit of 30% hydrogen is the majority end use case, followed by the transportation sector. For the minimum penetration scenario, transportation is overwhelmingly the largest use case, dominated by passenger FCVs for both scenarios.

There are several policy issues which need to be addressed in light of global and regional carbon reduction goals. The rapidly increasing energy system cost is a major issue, particularly for non-OECD nations who are rapidly modernizing. A significant part of these cost increases come from infrastructure costs for both hydrogen and CCS requirements, which are additional to current system costs. Secondly, the differentiation of carbon reduction targets causes an unforeseen issue whereby non-OECD nations with a more relaxed $CO_2$ reduction regime are able to produce hydrogen for OECD nations using fossil fuel sources, creating some perverse policy incentives which need to be addressed if CCS and hydrogen are to play a strong future role in meeting these targets. A potential solution to this issue is the improvement of the efficiency and cost of electrolysis technologies, coupled with further reductions in renewable energy costs in order to engender local production and consumption of hydrogen.

Although the level of hydrogen anticipated in this research is modest, the level of contribution in certain sectors, notably transportation, and toward carbon reduction is significant. In order to expand the role for hydrogen, exogenous stimuli are likely required, in the form of subsidies and preferential industrial treatment for related industries.

The incorporation of stakeholder engagement alongside global energy system modeling in the case study nation of Japan offers additional insights on the future deployment of passenger vehicles in a high hydrogen use scenario. Stakeholders are concerned about efficiency, environmental, cost and infrastructure accessibility when making future vehicle purchases. For a nation dependent on foreign imports for fuel, reducing $CO_2$ levels through the import of hydrogen is a likely future strategy for the Japanese government, able to reduce emissions and in tandem with future technological progression, meet reasonable cost levels.

The findings for Japan have relevance for other future hydrogen transport using regions identified in this study, among them North America, Western Europe, China and South East Asia.

As the model employed in this research is under constant development, some limitations are apparent, i.e. the model is not cognizant of all possible end-uses of hydrogen, particularly emerging solid oxide fuel cell (SOFC) deployment, or its use in steel reformation. Further, while the model is cognizant of several passenger vehicle types, plug-in hybrids are not considered at this stage.

## Acknowledgment

We would like to extend our thanks to Toyota Automotive Company for their kind support of our research and to Mr. Nakahara of the University of Tokyo for guidance regarding model settings and augmentation.

## References

[1] Rogelj, J. *et al.* Mitigation Pathways Compatible with 1.5°C in the Context of Sustainable Development. In: Global Warming of 1.5°C. An IPCC Special Report on the impacts of global warming of 1.5°C above pre-industrial levels and related global greenhouse gas emission pathways, in the context of strengthening the global response to the threat of climate change, sustainable development, and efforts to eradicate poverty. In Press. 2018.

[2] United Nations Framework Convention on Climate Change (UNFCCC). Nationally Determined Contributions (NDCs). 2017. <https://unfccc.int/process/the-paris-agreement/nationally-determined-contributions/ndc-registry>

[3] Höhne, N., Elzen, M., Weiss, M. Common but differentiated convergence (CDC): a new conceptual approach to long-term climate policy, Climate Policy, 6(2), 181-199. 2006.

[4] Peker, M., Kocaman, A. S., Kara, B. Y. Benefits of transmission switching and energy storage in power systems with high renewable energy penetration. Applied Energy, 228, 1182–1197. 2018.

[5] Haszeldine, R. S., Flude, S., Johnson, G., & Scott, V. Negative emissions technologies and carbon capture and storage to achieve the Paris Agreement commitments. Philosophical Transactions of the Royal Society A: Mathematical, Physical and Engineering Sciences, 376(2119). 2018.

[6] Anandarajah, G., McDowall, W., & Ekins, P. Decarbonising road transport with hydrogen and electricity: Long term global technology learning scenarios. International Journal of Hydrogen Energy, 38(8), 3419–3432. 2013.

[7] Krzyzanowski, D. A., Kypreos, S., & Barreto, L. Supporting hydrogen based transportation: Case studies with Global MARKAL Model. Computational Management Science, 5(3), 207–231. 2008.

[8] Michalski, J., Poltrum, M., & Bünger, U. The role of renewable fuel supply in the transport sector in a future decarbonized energy system. International Journal of Hydrogen Energy, In Press. 2018.

[9] Ghahramani M., Nazari-Heris M., Zare K., Mohammadi-ivatloo B. Optimal Energy and Reserve Management of the Electric Vehicles Aggregator in Electrical Energy Networks Considering Distributed Energy Sources and Demand Side Management. In: Ahmadian A., Mohammadi-ivatloo B., Elkamel A. (eds) Electric Vehicles in Energy Systems. Springer, Cham. 2020.

[10] Ahmed A, Al-Amin AQ, Ambrose AF, Saidur R. Hydrogen fuel and transport system: a sustainable and environmental future. International Journal of Hydrogen Energy, 41(3), 1369-80. 2016.

[11] Jayakumar, A., Chalmers, A., & Lie, T. T. Review of prospects for adoption of fuel cell electric vehicles in New Zealand. IET Electrical Systems in Transportation, 7(4), 259–266. 2017.

[12] Davis, K., & Hayes, J. G. Fuel cell vehicle energy management strategy based on the cost of ownership. IET Electrical Systems in Transportation, 9(4), 226–236. 2019.

[13] Din, T., & Hillmansen, S. Energy consumption and carbon dioxide emissions analysis for a concept design of a hydrogen hybrid railway vehicle. IET Electrical Systems in Transportation, 8(2), 112–121. 2018.

[14] Taji, M., Farsi, M., & Keshavarz, P. Real time optimization of steam reforming of methane in an industrial hydrogen plant. International Journal of Hydrogen Energy, 43(29), 13110–13121. 2018.

[15] Jones, D. R., Al-Masry, W. A., & Dunnill, C. W. Hydrogen-enriched natural gas as a domestic fuel: an analysis based on flash-back and blow-off limits for domestic natural gas appliances within the UK. Sustainable Energy & Fuels, 2(4), 710–723. 2018.

[16] Muradov N. Low to near-zero $CO_2$ production of hydrogen from fossil fuels: status and perspectives. International Journal of Hydrogen Energy, 42(20), 14058-88. 2017.

[17] McPherson, M., Johnson, N., & Strubegger, M. The role of electricity storage and hydrogen technologies in enabling global low-carbon energy transitions. Applied Energy, 216, 649–661. 2018.

[18] Awan, A. B., Zubair, M., Sidhu, G. A. S., Bhatti, A. R., & Abo-Khalil, A. G. Performance analysis of various hybrid renewable energy systems using battery, hydrogen, and pumped hydro-based storage units. International Journal of Energy Research, 1–26. 2018.

[19] Nazari-Heris, M., Mirzaei, M. A., Mohammadi-Ivatloo, B., Marzband, M., & Asadi, S. Economic-environmental effect of power to gas technology in coupled electricity and gas systems with price-responsive shiftable loads. Journal of Cleaner Production, 244, 118769. 2020.

[20] Caliskan H, Dincer I, Hepbasli A. Energy, exergy and sustainability analyses of hybrid renewable energy based hydrogen and electricity production and storage systems: modeling and case study. Applied Thermal Engineering, 61(2), 784-98. 2013.

[21] Hosoya, Y., Fujii, Y. Analysis of energy strategies to halve $CO_2$ emissions by the year 2050 with a regionally disaggregated world energy model. Energy Procedia 43, 5853-5860. 2011.

[22] Chapman, A., Itaoka, K., Farabi-Asl, H., Fujii, Y., Nakahara, M. Societal Penetration of Hydrogen into the Future Energy System: Impacts of Policy, Technology and Carbon Targets. International Journal of Hydrogen Energy, 45, 3883-98. 2020.

[23] IPCC. Climate Change 2014: Synthesis Report. Contribution of Working Groups I, II and III to the Fifth Assessment Report of the Intergovernmental Panel on Climate Change. IPCC, Geneva, Switzerland. 2014.

[24] Fraunhofer ISE. Current and Future Cost of Photovoltaics. Long-term Scenarios for Market Development, System Prices and LCOE of Utility-Scale PV Systems. Study on behalf of Agora Energiewende. 2015.

[25] Bloomberg NEF. 2Q 2019 Global PV Market Outlook. 2019.

[26] Ministry of Economy, Trade and Industry. Ministry efforts toward the uptake of EV and PHV. 2017.

[27] Ministry of Economy, Trade and Industry. EV and FCV Strategic Roadmap. <https://www.meti.go.jp/press/2018/03/20190312001/20190312001-2.pdf>. 2018.

[28] World Nuclear Association. Nuclear Power in the European Union. <http://www.world-nuclear.org/information-library/country-profiles/others/european-union.aspx>. 2017.

[29] Ministry of Economy, Trade and Industry. The Strategic Road Map for Hydrogen and Fuel Cells. 2019. <https://www.meti.go.jp/english/press/2019/pdf/0312_002b.pdf>

[30] Hydrogen Council. Hydrogen Scaling Up: A sustainable pathway for the global energy transition. 2017.

[31] Federal Highway Administration. National Household Travel Survey 2017. Average vehicle occupancy by vehicle type. 2017.

[32] Pollet, B., Kocha, S., Staffell, I. Current status of automotive fuel cells for sustainable transport. Current Opinion in Electrochemistry 16, 90-95. 2019.

[33] IEA. Hydrogen: Tracking Clean Energy Progress. 2018. <https://www.iea.org/tcep/energyintegration/hydrogen/>

[34] Ajanovic, A., Haas, R. Economic and Environmental Prospects for Battery Electric‐ and Fuel Cell Vehicles: A Review. Fuel Cells, 19(5), 515-529. 2019.